\documentclass{camera}
\usepackage{graphicx}

\begin{document}

\title{RECENT RESULTS FROM KASCADE-GRANDE}

\author{
K.-H.~KAMPERT$^{a}$,
W.D.~APEL$^{b}$, 
F.~BADEA$^{b}$,
K.~BEKK$^{b}$, 
A.~BERCUCI$^{c}$,
M.~BERTAINA$^{d}$,
J.~BL\"UMER$^{b,e}$,
H.~BOZDOG$^{b}$,
I.M.~BRANCUS$^{c}$,
M.~BR\"UGGEMANN$^{f}$,
P.~BUCHHOLZ$^{f}$,
A.~CHIAVASSA$^{d}$,
F.~COSSAVELLA$^{e}$, 
K.~DAUMILLER$^{b}$, 
F.~DI~PIERRO$^{d}$,
P.~DOLL$^{b}$, 
R.~ENGEL$^{b}$,
J.~ENGLER$^{b}$, 
P.L.~GHIA$^{g}$,
H.J.~GILS$^{b}$,
R.~GLASSTETTER$^{a}$, 
C.~GRUPEN$^{f}$,
A.~HAUNGS$^{b}$, 
D.~HECK$^{b}$, 
J.R.~H\"ORANDEL$^{e}$, 
T.~HUEGE$^{b}$, 
H.O.~KLAGES$^{b}$, 
Y.~KOLOTAEV$^{f}$,
H.J.~MATHES$^{b}$, 
H.J.~MAYER$^{b}$, 
C.~MEURER$^{b}$,
J.~MILKE$^{b}$, 
B.~MITRICA$^{c}$,
C.~MORELLO$^{g}$,
G.~NAVARRA$^{d}$,
S.~NEHLS$^{b}$,
R.~OBENLAND$^{b}$,
J.~OEHLSCHL\"AGER$^{b}$, 
S.~OSTAPCHENKO$^{b}$, 
S.~OVER$^{f}$,
M.~PETCU$^{c}$, 
T.~PIEROG$^{b}$, 
S.~PLEWNIA$^{b}$,
H.~REBEL$^{b}$, 
A.~RISSE$^{h}$, 
M.~ROTH$^{b}$, 
H.~SCHIELER$^{b}$, 
O.~SIMA$^{c}$, 
M.~ST\"UMPERT$^{e}$, 
G.~TOMA$^{c}$, 
G.C.~TRINCHERO$^{g}$,
H.~ULRICH$^{b}$,
J.~VAN~BUREN$^{b}$,
W.~WALKOWIAK$^{f}$,
A.~WEINDL$^{b}$,
J.~WOCHELE$^{b}$, 
J.~ZABIEROWSKI$^{h}$ \and
D.~ZIMMERMANN$^{f}$\\[2ex]
{\bf KASCADE-Grande Collaboration}
}

\organization{
$^{a}$ Fachbereich C - Physik, Univ.\ Wuppertal, 42119 Wuppertal, Germany \\
$^{b}$ IK, Forschungszentrum Karlsruhe, 76021~Karlsruhe, Germany\\
$^{c}$ Nat. Inst. of Physics and Nuclear Eng., 7690~Bucharest, Romania\\
$^{d}$ Dipartimento di Fisica Generale dell'Universit{\`a},
10125 Torino, Italy\\
$^{e}$ IEKP, Universit\"at Karlsruhe, 76021 Karlsruhe, Germany,\\
$^{f}$ Fachbereich Physik, Universit\"at Siegen, 57068 Siegen, 
Germany \\
$^{g}$ Istituto di Fisica dello Spazio Interplanetario, INAF, 
10133 Torino, Italy \\
%
%
% $^{h}$ Soltan Institute for Nuclear Studies, 90950~Lodz, 
% Poland\\[1.5ex]
% %
% $^{1}$ on leave of absence from Nat.\ Inst.\ of Phys.\ and 
% Nucl.\ Engineering, Bucharest, Romania\\
% %
% $^{2}$ on leave of absence from Moscow State University, 
% 119899~Moscow, Russia\\
% %
}

\maketitle

\newpage
\begin{abstract}
KASCADE-Grande is a new extensive air shower experiment
co-located to the KASCADE site at Forschungszentrum Karlsruhe.
The multi-detector system allows to investigate the energy
spectrum, composition, and anisotropies of cosmic rays with
unprecedented prevision in the energy range from
$10^{14}$-$10^{18}$~eV. The primary goals besides investigating
the origin of the knee at $E \simeq 3 \cdot 10^{15}$~eV, are to
verify the existence of the second knee at $E \sim 10^{17}$~eV
and to measure the composition in the expected transition region 
of galactic to extragalactic cosmic rays.  The performance of the
apparatus and shower reconstruction methods will be discussed on
the basis of detailed Monte Carlo simulations and first data.
First results based on slightly more than a year of data taking
are presented.
\end{abstract}

\section{Introduction}

\begin{figure*}[t]
\centering
\vspace*{0.1cm}
\includegraphics[width=100mm]{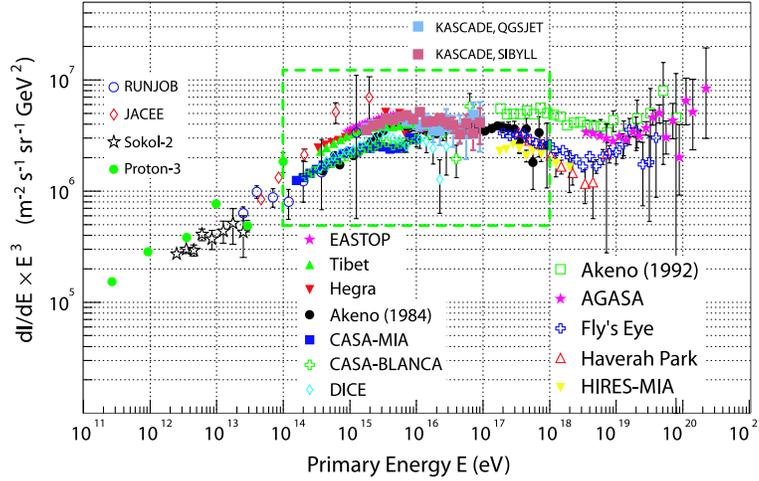}
\caption[xx]{The all-particle CR energy spectrum weighted by
$E^{3}$ showing the knee at $3\cdot10^{15}$~eV, a possible second
knee at $\sim 10^{17}$~eV, and the ankle at about
$3\cdot10^{18}$~eV. The energy range covered by KASCADE-Grande is
indicated by the dashed box.}
\label{fig:knee}
\end{figure*}

Despite intense research over several decades and measuring an
impressive set of cosmic ray (CR) data, the origin of non-solar
CRs remains unknown.  As an example about the available data,
Fig.\,\ref{fig:knee} shows the so called {\em all-particle}
spectrum.  It has become standard practice to present the
spectral data as the flux $dJ/dE$ times a power of the energy
$E$, so that the visibility of spectral features is emphasized.
The most prominent feature in the spectrum is the {\em knee} at
about $3\cdot10^{15}$~eV where the index $\gamma$ of the power
law spectrum $dJ/dE \propto E^{-\gamma}$ changes from $\gamma
\simeq 2.75$ to $\simeq 3.1$.  The origin of this phenomenon is
still unknown, but it is generally attributed to either the
limiting energy of galactic CR accelerators and/or to the onset
of strong diffusion losses of particles out of our Galaxy.  Also,
particle physics explanations have been put forward, either
assuming the onset of a new interaction mechanism above the knee
energy so that part of the CR-energy remains invisible to
experiments or by considering propagation effects between the
sources and the solar system.  Generally, supernova remnants
(SNR) are conceived to be the major sources for CRs up to the
knee with the acceleration taking place by the first order Fermi
mechanism.  Based on simple dimensional arguments, the maximum
energy would then be given by $E_{\rm max} \propto Z \cdot R
\cdot B$, with $Z$ being the charge of the CR particle and $R$
and $B$ being the size and magnetic field of the source,
respectively.  As a consequence, the knee position for different
CR primaries would be expected to scale with their atomic number
$Z$.  Similarly, diffusion losses out of the Galaxy would result
in a $Z$ dependence of the knee position, since magnetic
confinement is the underlying process.  For a recent review on
astrophysical models, the reader is referred to \cite{Hillas05}
and references therein.  Particle physics models generally
predict a scaling of the knee energy with the atomic mass $A$,
since the reaction mechanism at high energies is governed by the
energy per nucleon rather than by the total energy of the CR
particles.  More recently, the cannonball (CB) model was
suggested to explain the origin of all non-solar CRs
\cite{Dar06}.  It predicts a scaling of the knee position with
$A$ rather than with $Z$, because of deflections of CRs at the
relativistically moving CBs.  In any of the astrophysical models,
light particles drop out of the acceleration and/or confinement
region first -- this is what should cause the knee -- so that the
composition is expected to change from light to heavy when
crossing the knee energy.

A na\"{i}ve extrapolation of this idea suggests that the putative
second knee would mark the position of the Fe-knee.  No consensus
on a preferred accelerator site or mechanism exists for the
energies between the (second) knee and ankle.  It has long being
argued that CRs above the ankle are of extragalactic origin, as
they cannot be isotropized by the Galactic magnetic fields, but
their arrival directions are surprisingly isotropic.  At the
highest energies, above $5 \cdot 10^{19}$~eV, the
GZK\footnote{Greisen, and Zatsepin \& Kuzmin} cut-off is
expected.  Because of the low flux involved, such energies cannot
be measured with KASCADE-Grande.  This is the domain of the
Pierre Auger Observatory \cite{Auger04}.

\section{The KASCADE-Grande Experiment}

The multi-detector system KASCADE-Grande (KArlsruhe Shower Core
and Array DEtector and Grande array) is optimized for the energy
range $10^{14}$ - $10^{18}$~eV and addresses the questions of the
previous section by measuring as much as possible from each
single air-shower event \cite{Kampert03,Navarra04}.  The main
detector components of KASCADE-Grande are the KASCADE array, the
Grande array, the Central Detector, and the Muon Tracking
Detector.

\begin{figure}[bh]
\begin{minipage}{14pc}
\centering
\includegraphics[width=13pc]{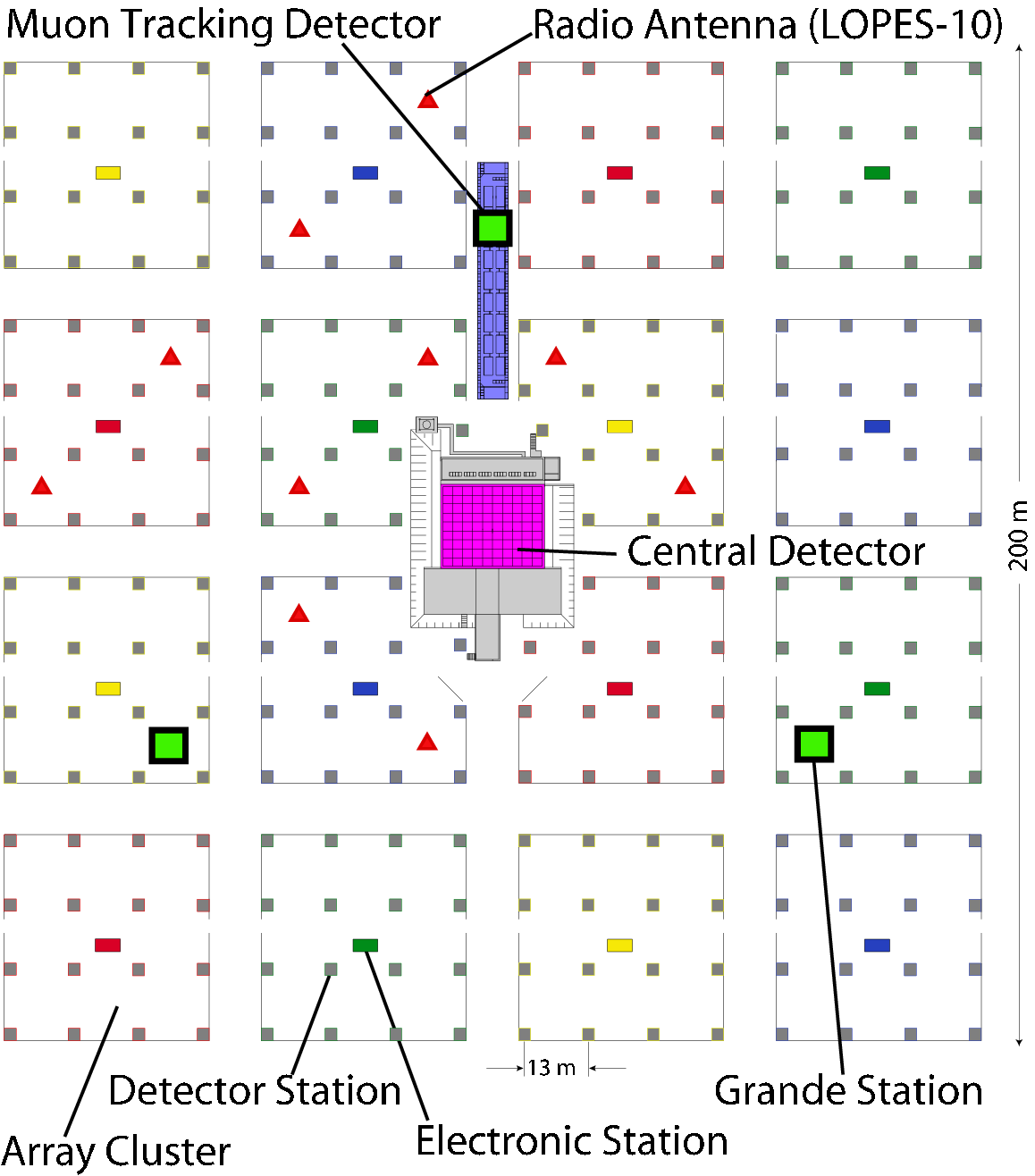}
\caption{\label{fig:kascade-array}
The main detector components of the KASCADE experiment: 
(the 16 clusters of) Field Array, Muon Tracking Detector and 
Central Detector. The location of 10 radio antennas is also 
displayed, as well as three stations of the Grande array.}
\end{minipage}\hspace{2pc}
\begin{minipage}{14pc}
\includegraphics[width=13pc]{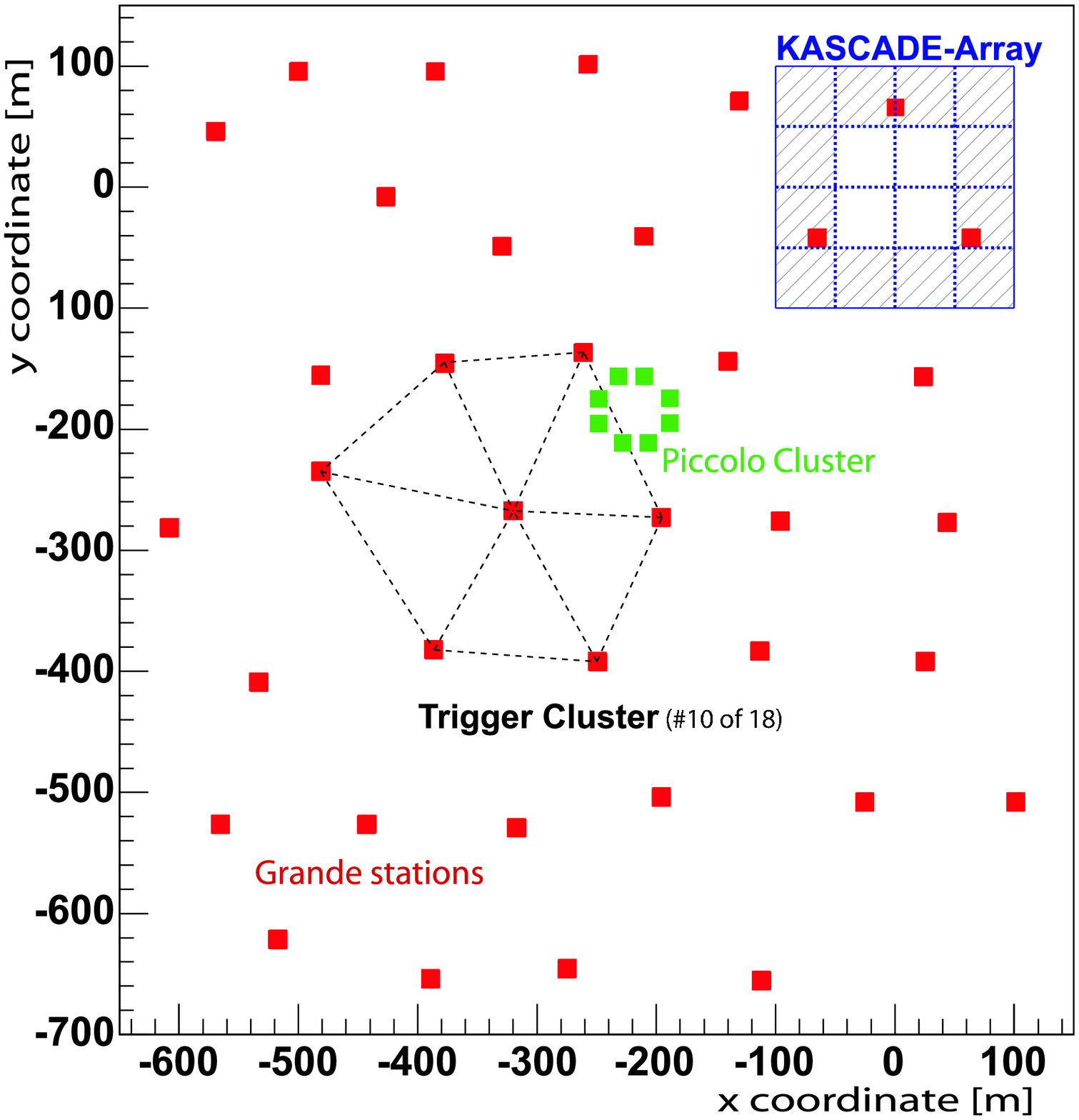}
\caption{\label{fig:grande-array}Layout of the KASCADE-Grande
experiment with its 37 Grande and 8 Piccolo stations.  One of the
18 Grande trigger hexagons is also shown.}
\end{minipage} 
\end{figure}

The KASCADE array (Fig.\ \ref{fig:kascade-array}) measures the
total electron and muon numbers ($E_\mu>230\,$MeV) of the shower
by using an array of 252 detector stations containing both
shielded and unshielded detectors located on a grid of $200
\times 200\,$m$^2$ ~\cite{kas}.  The excellent timing resolution
of these detectors allows also investigations of the arrival
directions of the showers in searching for large scale
anisotropies \cite{Gmaier2} as well as for cosmic ray point
sources \cite{Gmaier1}.  The KASCADE array is optimized to
measure extensive air showers (EAS) in the energy range of
$10^{14}$ eV to $8 \cdot 10^{16}$ eV and has provided for the
first time energy spectra of CR mass groups \cite{Ulrich05}.  The
Muon Tracking Detector ($128\,$m$^2$) measures the incidence
angles of muons ($E_\mu > 800\,$MeV) relative to the shower
arrival direction \cite{Doll02}.  These measurements provide a
sensitivity to the longitudinal development of the showers.

Muons with energies above 2.4 GeV are measured in the
300\,m$^{2}$ large central detector by a multiwire proportional
chambers (MWPC) and limited streamer tubes (LST).

The 37 stations of the Grande Array (Fig.~\ref{fig:grande-array})
extend the cosmic ray measurements up to primary energies of
\mbox{1 EeV}.  The Grande stations, \mbox{10 m$^2$} of plastic
scintillator detectors each, are positioned at an average mutual
distance of approx.\ \mbox{130 m} covering a total area of
\mbox{$\sim$ 0.5 km$^2$}.  There are 16 scintillator sheets in a
station read out by 16 high gain photomultipliers; 4 of the
scintillators are read out also by 4 low gain PMs.  The covered
dynamic range is up to \mbox{3000 mips/m$^2$}.  A trigger signal
is build when 7 stations in a hexagon (trigger cluster, see
Fig.~\ref{fig:grande-array}) are fired.  Therefore, the Grande
array consists of 18 hexagons with a total trigger rate of
\mbox{0.5 Hz}.

Additionally to the Grande Array a compact array, named Piccolo,
has been build in order to provide a fast trigger to KASCADE
ensuring joint measurements for showers with cores located far
from the KASCADE array.  The Piccolo array consists of 8 stations
with \mbox{11 m$^2$} plastic scintillator each, distributed over
an area of \mbox{360 m$^2$}.  One station contains 12 plastic
scintillators organized in 6 modules; 3 modules form a so-called
electronic station providing ADC and TDC signals.  A Piccolo
trigger is built and sent to KASCADE and Grande when at least 7
out of the 48 modules of Piccolo are fired.  Such a logical
condition leads to a trigger rate of \mbox{0.3 Hz}.

\begin{figure*}[t]
\centering
%\vspace*{0.1cm}
\includegraphics[width=70mm]{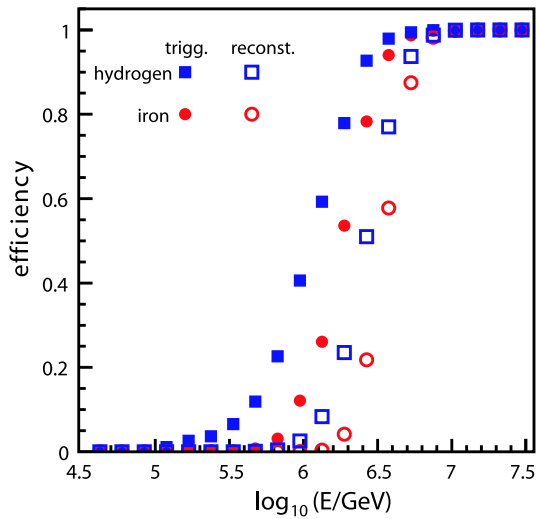}
\caption[xx]{Trigger and reconstruction efficiency for Hydrogen
and Iron induced EAS (zenith angles $< 18^{\circ}$) as function
of primary energy.
\label{fig:efficieny}}
\end{figure*}

The trigger and reconstruction efficiency of KASCADE-Grande is
depicted in Fig.\,\ref{fig:efficieny} \cite{vanBuren06}.  Full
efficiency is reached already at around $10^{16}$~eV so that a
significant region of overlap for cross-checks is attained with
KASCADE.

Presently, a self-triggering, dead-time free FADC-based DAQ
system is being implemented in order to record the full time
evolution of energy deposits in the Grande stations at an
effective sampling rate of \mbox{250 MHz} and high resolution of
12 bits in two gain ranges~\cite{Walkowiak05}.  This system runs
in parallel to the present data acquisition system and will allow
for an intrinsic electron-muon separation by the time traces
measured in the Grande stations.

The whole KASCADE-Grande setup is read out if a certain
multiplicity of the KASCADE, Piccolo, or the Grande array
detector stations is firing, leading to a total trigger rate of
$\approx 4.5\,$Hz.

The redundant information of the showers measured by the Central
Detector and the Muon Tracking Detector is predominantly being
used for tests and improvements of the hadronic interaction
models underlying the analyses~\cite{isv04}.

For the calibration of the radio signal emitted by the air shower
in the atmosphere an array of first 10 and meanwhile 30 dipole
antennas (LOPES) is set up on the site of the KASCADE-Grande
experiment~\cite{lopes-nature,lopes}.

\section{Shower Reconstruction}

The key observables for reconstruction the primary energy and
mass of CRs are given by the electron and muon numbers at ground.
The KASCADE detector stations allow for a separation of electrons
and muons because of a 20 radiation lengths shielding used above
the muon counters.  In case of the KASCADE-Grande stations such
shielding was not available, so that at present only numbers of
{\em charged particles} can be obtained.  This will change in the
future with the availability of the FADC readout system
\cite{Walkowiak05}.  For the time being, the muon numbers are
obtained relatively locally from the KASCADE array.  The shower
geometry of EAS landing within the fiducial area of Grande are
reconstructed by the particle densities and timing information of
the Grande stations and the muons are extracted from the muon
densities measured in KASCADE. An important step in such an
analysis is the determination of the lateral density
distributions (LDF) of electrons and muons.  In hadronic induced
air showers especially at large core distances a slightly modified
NKG\footnote{NKG = Nishimura, Kamata, Greisen}-function is used 
for the electrons:
\begin{equation}
    \rho_{e} = N_e \cdot C(s) \cdot \left(
\frac{r}{r_0}\right)^{s-\alpha} \left(
1+\frac{r}{r_0}\right)^{s-\beta},
\label{eq:electron-ldf}
\end{equation}
with the normalization factor $C(s)=\Gamma(\beta-s) /
\Gamma(s-\alpha+2) / \Gamma(\alpha+\beta-2s-2) / (2 \pi r_0^2)$,
the shower size $N_e$, and the so-called shower age $s$.
Performing CORSIKA \cite{corsika} simulations and taking into
account the experimental response of the Grande array, the
parameters $\alpha=1.5$, $\beta=3.6$, and $r_0=40$~m were found
to describe the lateral distribution function best
\cite{Glasstetter-05}.  The same set of parameters also provides
a better description of the high precision KASCADE data, as was
pointed out recently \cite{Mayer06}.  Such a deviation is not
surprising, since the traditional NKG-formula with $\alpha=2$,
$\beta=4.5$ and $r_0=89$~m has been derived analytically for the
case of pure $e/\gamma$-showers.

To describe the average arrival time $\bar{t}$ and the time
spread $\sigma_t$ of the shower electrons, a Linsley-function has
been adapted to the time distributions of pure CORSIKA
simulations \cite{Glasstetter-05}:
 \[ \bar{t} = 2.43 \cdot(1 + \frac{r}{30\mbox{ m}})^{1.55} \;\;\;\; \mbox{and} \,\,\,\,
    \sigma_t = 1.43 \cdot (1 + \frac{r}{30\mbox{ m}})^{1.39}  \]
The parameters were found to dependend only weakly on the primary
particle properties.  To first order approximation the measured
arrival time of the first out $N$ of particles inside a detector
is given by $\bar{t}_{\rm 1.\,of\,N} = \bar{t}/\sqrt{N}$.

\begin{figure}[t]
\begin{minipage}{13.9pc}
\centering
\includegraphics[width=60mm]{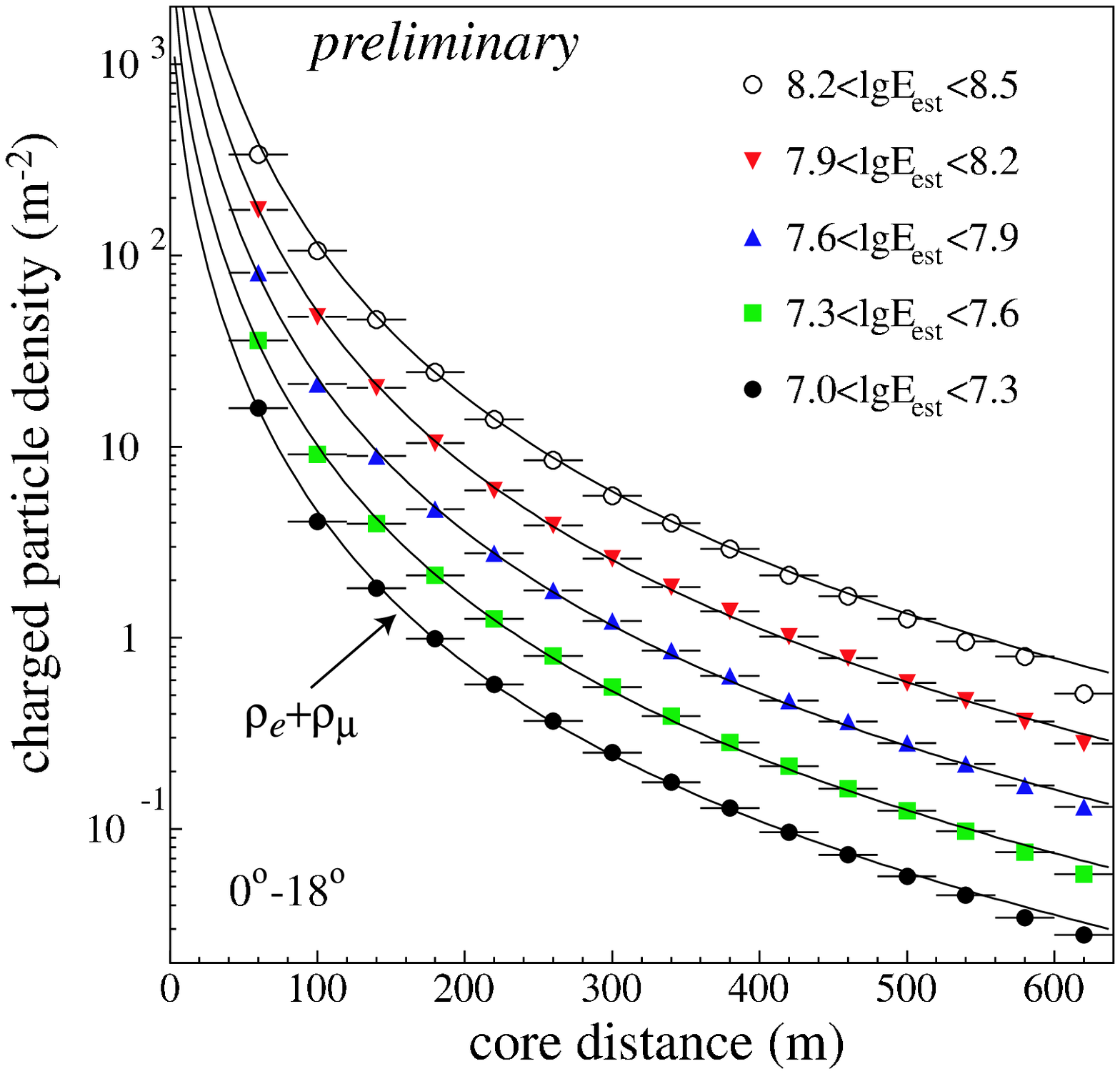}
\caption[xx]{Measured lateral distribution of the charged 
particle component for five estimated energy
intervals \cite{Glasstetter-05}. The lines represent the sum of 
the LDFs of eqn.\ \ref{eq:electron-ldf} and \ref{eq:muon-ldf} 
fitted to the data.\label{fig:charged-ldf}}
\end{minipage}\hspace{2pc}
\begin{minipage}{13.9pc}
\includegraphics[width=14pc]{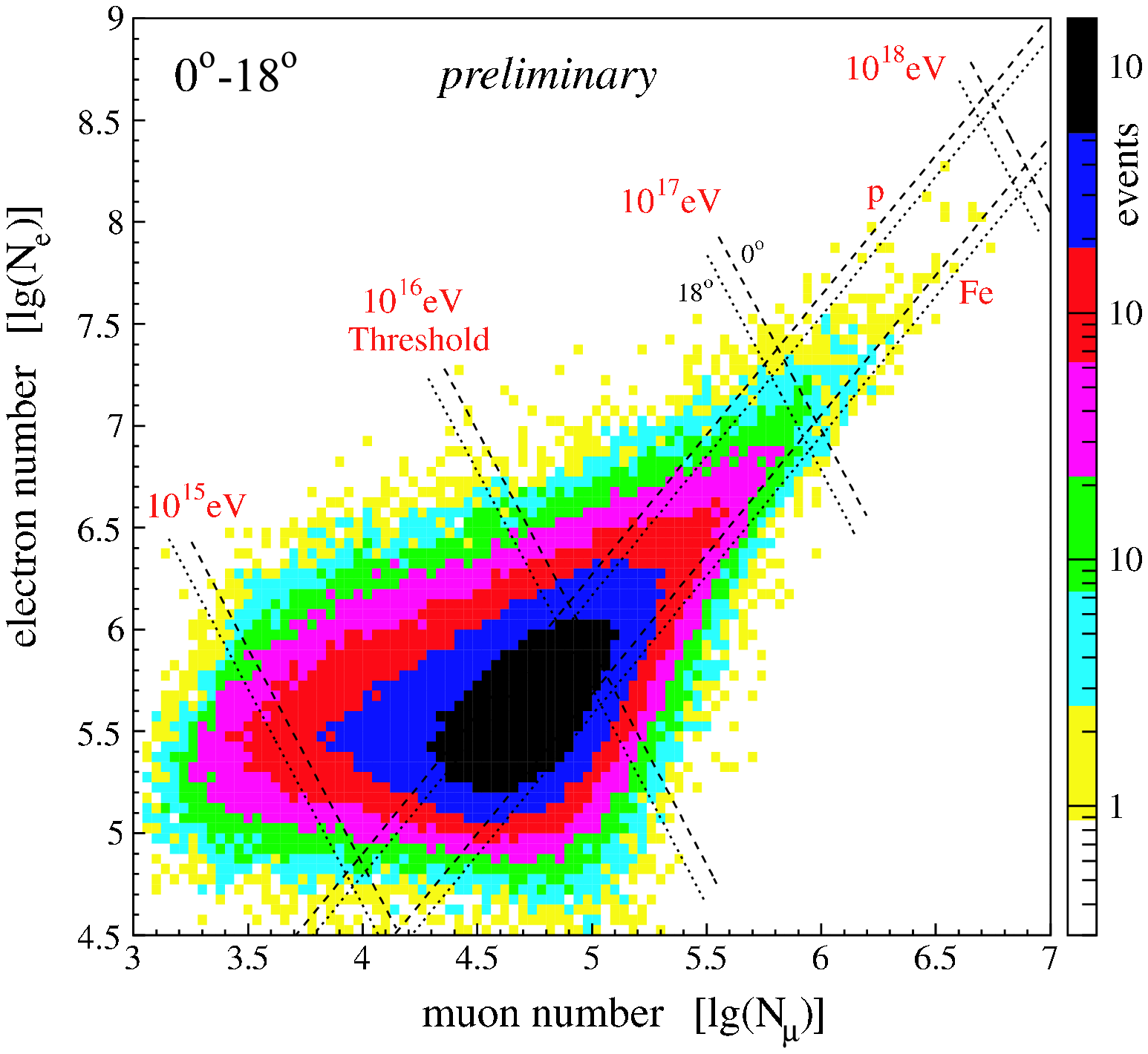}
\caption{\label{fig:ne-nmu}Reconstructed electron and muon
numbers from KASCADE-Grande.  The dashed line pairs indicate
average lines of constant energy and mass derived from CORSIKA
simulations for the extreme zenith angles.}
\end{minipage} 
\end{figure}

Since the functions above are coupled via the particle number and
the core distance in shower disc coordinates, they are fitted
simultaneously to the data in a combined
neg.-log-likelihood/$\chi^2$ minimization.  For the calculation
of the expected particle density in a Grande station a
contribution from the previously reconstructed muon lateral
distribution function is taken into account.  Thus, the free 7
parameters of the global fit are the core position and shower
direction (incl.\ a global time offset), as well as the electron
number ($E_{kin}>3$ MeV) and shower age. For the muon LDF a 
parametrization similar to \cite{lagutin} has been used:
\begin{equation}
    \rho_{\mu} = N_\mu \cdot \frac{0.28}{r_{0}^{2}}
    \left( \frac{r}{r_0}\right)^{p1} \cdot
    \left( 1+\frac{r}{r_0}\right)^{p2} \cdot
    \left( 1 + \left( \frac{r}{10 \cdot r_{0}} \right)^{2} \right)^{p3}
\label{eq:muon-ldf}
\end{equation}
with $r_{0}=320$m and $p1=-0.69$, $p2=-2.39$, $p3=-1.0$ when
averaging the fits to the LDF in the energy range $10^{16}$\,eV -
$10^{17}$\,eV \cite{vanBuren05,vanBuren06}. The result of 
$\rho_{e}+\rho_{\mu}$ fitted to the data by eqns.\ 
\ref{eq:electron-ldf} and \ref{eq:muon-ldf} is shown Fig.\ 
\ref{fig:charged-ldf}.

The reconstruction quality has been studied extensively in
Ref.~\cite{Glasstetter-05}.  Above a threshold of 10$^6$
electrons corresponding to 100\% trigger efficiency the shower
core and direction resolution is reconstructed to better than
12~m resp.\ $0.6^\circ$ and they improve to better than 8~m
resp.\ $0.4^\circ$ at $10^{7}$ electrons.  At the same time, the
statistical uncertainties of the electron and muon numbers are
below $\Delta \log N_{e,\mu} = 10 \%$ and 20\,\%, respectively.
Systematic deviations are always well below the statistical
accuracy of the experiment.

The resulting distribution of electron and muon numbers is
presented in Fig.\ \ref{fig:ne-nmu} for zenith angles below
$18^{\circ}$.  Even though full efficiency is only obtained at
$10^{16}$~eV, most events are recorded at lower energies which is
due to the steeply falling energy spectrum of CRs.  The dashed
line pairs indicate energy and mass ranges as derived from
CORSIKA simulations for zenith angles of $0^{\circ}$ and
$18^{\circ}$ and using the relation presented in the next
section.

\section{First Data}

The data presented in the following were taken from March 2004 to
March 2006.  Due to calibrations and tests at KASCADE-Grande
still going on, this corresponds to an effective data taking time
of 363 days when accepting only periods where both KASCADE and
KASCADE-Grande were operating with all components and without
failures.  During this period approximately 15.1 million events
have been registered by the Grande array.  In order to be
considered in this preliminary analysis, at least 20 Grande
stations with measured energy deposits were requested with the
reconstructed shower core to be within a circle of 250 m radius
around the center of the Grande array.  The reconstructed age
parameter was restricted to $0.4 < s < 1.4$.  These cuts are
applied to reduce the effect of misreconstructed cores at the
border of the array, especially from showers being originally
outside the Grande array and to ensure good electron size
reconstruction also at the trigger threshold.  It has been
verified by simulations that these cuts do not result in an
increase of the energy threshold of the experiment.  Still,
100\,\% efficiency is maintained both for proton and iron
primaries at $E \geq 10^{16}$\,eV which provides a large region
of overlap with the KASCADE data.  However, the total number of
events in the sample is reduced by these cuts to $6.50 \cdot
10^{4}$ and $4.67 \cdot 10^{4}$ for zenith angles
$0^{\circ}$-$18^{\circ}$ and $18^{\circ}$-$25^{\circ}$,
respectively.

\begin{figure*}[t]
\centering
\includegraphics[width=80mm]{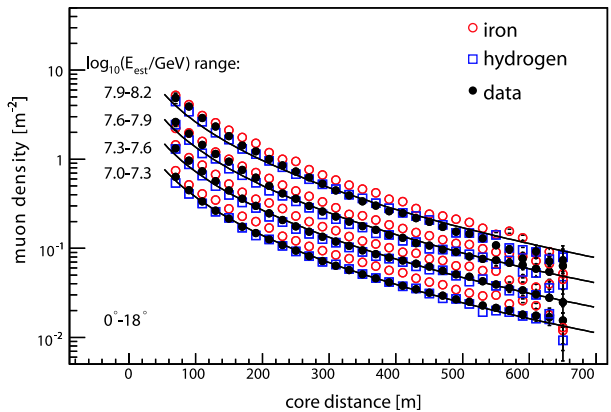}
\caption[xx]{Measured and simulated muon density distribution for
zenith angles $0^{\circ}$-$18^{\circ}$ and four estimated energy
intervals \cite{vanBuren06}.  The lines represent the LDF of eqn.\
\ref{eq:muon-ldf}.
\label{fig:muon-ldf}}
\end{figure*}

Figure \ref{fig:muon-ldf} presents the measured muon LDFs for
different primary energies in comparison to CORSIKA simulations
for proton and iron primaries.  The energy has been estimated
from the electron and muon numbers by
 \[ \log_{10}(E_{\rm est}/{\rm GeV}) =
    0.313 \cdot \log_{10} N_{e} + 0.666 \cdot \log_{10} N_{\mu} +
    1.24 / cos \theta + 0.580 . \]
Generally, the agreement between data and MC is very good with
data being always contained between p and Fe simulations.  The
LDF (eq.\ \ref{eq:muon-ldf}) describes the data reasonable well up
to $\sim 10^{17}$~eV but appears somewhat too flat at higher
energies.

\begin{figure*}[t]
\centering
\includegraphics[width=80mm]{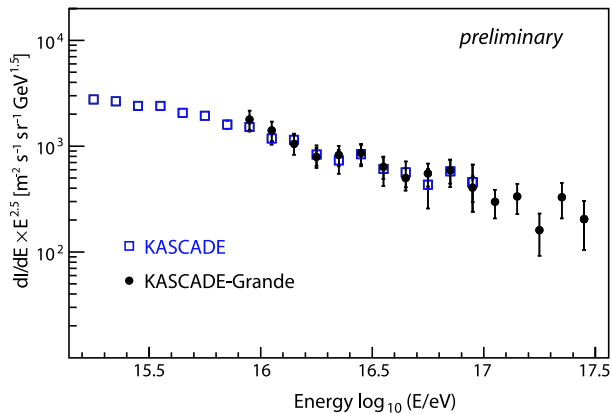}
\caption[xx]{Unfolded all-particle energy spectrum of
KASCADE-Grande compared to data of KASCADE from Ref.\
\cite{Ulrich05} for QGSJet based CORSIKA simulations and for
zenith angles $0^{\circ} \leq \theta \leq 18^{\circ}$.  The data
correspond to one year of KASCADE-Grande and are of preliminary
nature.
\label{fig:e-spectrum}}
\end{figure*}

In Ref.\ \cite{Ulrich05} we have emphasized the importance of
unfolding algorithms for the reconstruction of the energy and
mass of cosmic rays.  This is because of the large fluctuations
of shower observables affecting any measurement in the presence
of steeply falling spectra.  Because of the preliminary nature of
the data, we have not yet attempted a 2-dimensional unfolding of
the electron and muon numbers such as in Ref.\,\cite{Ulrich05}.
however, since the primary energy is mostly determined by the
muon number with only a weak dependence on the primary mass, an
unfolding of the muon size spectrum can be used as a first step
to determine the CR energy spectrum at energies beyond the
KASCADE range.  Again, the iterative Gold algorithm was employed.
As an example, Fig.\,\ref{fig:e-spectrum} shows the results of
unfolding for the zenith angle range $0^{\circ} \leq \theta \leq
18^{\circ}$.  The error bars of the total spectrum indicate the
total statistical error as squared sum of Poissonian statistics
and those resulting from the unfolding.  The results are compared
with previous ones from KASCADE and are also based on the QGSjet
assumption.  One observes a very good overlap in the energy range
of $10^{16}$~eV to $10^{17}$~eV. The KASCADE-Grande energy
spectrum reaches up to an energy of $3 \cdot 10^{17}$~eV. The
statistics is still too poor to draw any conclusion about a
second knee.  The quality of the solution can be characterized by
the $\chi^{2}$ value, which is obtained by folding the solution
with the response matrix and comparing the resulting vector with
the original data vector.  For the first and second zenith angle
ranges one obtains $\chi^{2}/{\rm dof}$ = 2.59 and 2.25
respectively, indicating the still imperfect description of the
data.  This is expected to improve with the presently applied new
calibration of the Grande stations.

\section{Summary and Conclusions}

KASCADE-Grande has started to take data.  All components are
working stable and perform well.  The presently installed FADC
system will extend its capabilities and provide detailed
information about the longitudinal profile of the shower disk and
allow for additional electron - muon separation.  The LDFs of
both electrons and muons are found to agree well with EAS
simulations.  From their obtained electron and muon numbers a
first estimate of the primary energy spectrum could be derived.
Within the overlap region with KASCADE, a good agreement is
found.  However, the statistics beyond $3 \cdot 10^{17}$\,eV is
still to poor to allow conclusions about the second knee.  This
will change in the near future with more data becoming available.
KASCADE-Grande keeps the multi-detector concept for tuning
different interaction models at primary energies up to
\mbox{$10^{18}$ eV}.  It also provides a perfect environment
for detecting radio emission in extensive air showers, which is
the aim of the LOPES project~\cite{lopes}.

\medskip 
\noindent {\bf Acknowledgments} \\
KHK wishes to thank the organizers for their invitation to the
Vulcano workshop conducted in a very pleasant and fruitful
atmosphere.  KASCADE-Grande is supported by the Ministry for
Research and Education of Germany, the INFN of Italy, the Polish
State Committee for Scientific Research (KBN grant for 2004-06)
and the Romanian National Academy for Science, Research and
Technology.

%\bigskip
% \bigskip
% \noindent {\bf DISCUSSION}
% 
% \bigskip
% \noindent {\bf TERESA MONTARULI:} I am not so happy about the 
% fact that most of the CR experiments use QGSJet 01 even though 
% it generates too few muons. We have very precise measurement of 
% us from AMANDA at high energy and from L3 below 2 TeV that 
% clearly show that the muon normalization predicted by QGSJet 01 
% is low by about 30\,\%. As a matter of fact, there exists QGSJet 
% 02 that already predicts 15\,\% more muons. I think, these 
% experiments should try to update the hadronic models.
% 
% \bigskip \noindent {\bf KARL-HEINZ KAMPERT:} Yes, what you say
% about AMANDA and L3 is correct, i.e.\ QGSjet 01 predicts too few
% muons at energies of the TeV scale.  However, KASCADE is
% sensitive to muons at the GeV scale where GQSJet 01 performs
% quite well, but where SIBYLL 2.1 predicts too few muons.
% Moreover, as you know one cannot easily change one parameter in
% the models so that muons are described satisfactory and all the
% rest would remain unaffected.  Any change of one parameter causes
% complex changes and requires detailed comparisons to existing
% accelerator and CR data.

\end{document}